\documentclass[a4paper,twocolumn,english,amsmath,amssymb,nofootinbib,showpacs,preprintnumbers]{revtex4}
\usepackage[T1]{fontenc}
\usepackage[latin1]{inputenc}
\usepackage{amssymb}

\makeatletter

\usepackage{graphicx}
\usepackage{amssymb}

\usepackage{ae, aecompl}

\usepackage{dcolumn}
\usepackage{bm}
\usepackage{graphics}
\usepackage{epsfig}

\newcommand{\Mplr}{M_{\textrm{Pl}}}
\newcommand{\kl}[1]{{\textstyle #1}}
\newcommand{\BmL}{$B$$-$$L$}

\usepackage{babel}

\makeatother
\begin{document}

\preprint{TUM-HEP-603/05}

\title{Leptonic Dark Energy and Baryogenesis}

\author{Florian Bauer}

\email{fbauer@ph.tum.de}

\author{Marc-Thomas Eisele}

\email{eisele@ph.tum.de}

\author{Mathias Garny}

\email{mgarny@ph.tum.de}

\affiliation{Physik Department T30d, Technische Universit\"{a}t M\"{u}nchen\\
 James-Franck-Stra\ss{}e, 85748 Garching, Germany}

\begin{abstract}
We consider a baryogenesis scenario, where the difference of baryon~($B$)
and lepton~($L$) number is conserved in such a way that the $B$$-$$L$
asymmetry in the standard model sector is compensated by an asymmetry
of opposite sign stored in the dark energy sector. Therefore, we introduce
a toy-model in which a complex quintessence field carries a $B$$-$$L$
asymmetry at late times. We determine the produced baryon asymmetry
in the visible sector for a large range of initial conditions and
find it easy to achieve a value of the observed order of magnitude.
While the size of the produced baryon asymmetry depends on details
of the underlying inflationary model, it turns out to be independent
of the reheating temperature in many cases. We also discuss possible
sources of instability like the formation of Q-Balls. 
\end{abstract}

\pacs{11.30.Fs, 98.80.Cq}

\maketitle

\section{Introduction}

One of the most intriguing questions in cosmology and particle physics
concerns the origin of the observed asymmetry between baryons and
anti-baryons~\cite{Spergel:2003cb}. If we assume that our universe
underwent an inflationary phase, any initial asymmetry has been diluted
during inflation and some baryogenesis process must have generated
the observed asymmetry afterwards. Frequently it is stated that any
baryogenesis process has to fulfill the three Sakharov conditions:
a) $B$-violation, b) $C$- and $CP$-violation, and c) departure
from thermal equilibrium~\cite{Sakharov:1967dj}. Since only \BmL{}
is conserved in the standard model (SM), most baryogenesis models
introduce new $B$ and/or $L$ violating interactions at some high
energy scale. Typical scenarios of this kind include heavy particle
decays or make use of scalar field dynamics. An example for the latter
is the Affleck-Dine mechanism~\cite{Affleck:1984fy}, where complex
scalar superpartners of fermionic fields acquire baryonic and/or leptonic
charges due to \BmL{} violating interactions in their potentials.
These asymmetries are later on transferred to the SM fermions. However,
it is also possible to create a baryon asymmetry in the SM particle
sector within a $B$-conserving theory. All one has to do is to store
a $B$ (or \BmL{}) asymmetry of opposite sign in some hidden sector,
see e.g.\ Ref.~\cite{Dodelson:1989ii}. The case that this sector
is dark matter and that it is completely made up of antibaryons has
first been been considered in Ref.~\cite{Barr:1990ca}, where the
ratio of the energy density of SM particles and dark matter leads
to a prediction for the masses of the constituents of dark matter.
Another example for a \BmL{} conserving baryogenesis scenario is
found in Ref.~\cite{Dick:1999je}. 

Recently, another topic in cosmology has also drawn a lot of attention
since one has found evidence~\mbox{\cite{Tonry:2003zg,Knop:2003iy,Spergel:2003cb,Tegmark:2003ud,Boughn:2004zm}}
that the expansion of the universe is accelerating. One origin of
this acceleration might be some unknown energy form called dark energy~(DE),
see Refs.~\cite{Peebles:2002gy,Padmanabhan:2002ji,Sahni:2004ai,Padmanabhan:2005cw}
for some recent reviews. Quintessence, a homogeneous scalar field
rolling down a suitable chosen potential is a popular candidate for
DE~\cite{Wetterich:1987fm,Ratra:1987rm}. In most quintessence models
the scalar field is real-valued, however, the variant of a complex
scalar has also been considered \cite{Gu:2001tr,Boyle:2001du}. The
major difference results from the possibility to assign a global conserved~$U(1)$
quantum number to the complex scalar field, i.e.\ it can carry a
charge. Here, a possible connection to baryogenesis emerges. 

Since scalar fields could be responsible for the observed baryon asymmetry
as well as the acceleration of the universe, it seems natural to consider
the possibility that a scalar field could be connected to both phenomena.
While Ref.~\cite{Boyle:2001du} mentions several possible realizations
for such an idea, including the one of hiding anti-baryons in the
vacuum in a baryon-symmetric baryogenesis model, this paper introduces
a first model (to our knowledge) that draws a direct connection between
the charge of a complex field responsible for dynamical DE and the observed
baryon asymmetry in our part of the universe. For various other possible
connections of baryogenesis and a DE scalar field or the vacuum see also Refs. \cite{Li:2001st,DeFelice:2002ir,Li:2002wd,Yamaguchi:2002vw,Bi:2003yr,Gu:2003er,Volovik:2003hn}. 

In the presented scenario \BmL{} is conserved and a corresponding
asymmetry is stored in the DE sector compensating for the observed
baryon asymmetry and possible other asymmetries in hidden sectors.
Therefore we introduce a simple toy-model, in which an additional
scalar field mediates between a complex quintessence field (carrying
lepton number) and the fermionic sector. At early times relative phase
differences in the initial conditions will lead to lepton asymmetries
in both condensates. The additional field eventually decays and transfers
its asymmetry to the fermionic sector, where sphalerons \cite{Klinkhamer:1984di,Kuzmin:1985mm}
create a baryon asymmetry. The new field also acquires a huge mass
due to the large vacuum expectation value (VEV) of the quintessence
field and hereby effectively decouples the fermionic and the DE sector. 

The paper is organized as follows: in Sec.~\ref{sec:Scenario} we
define the Lagrangian of the model and discuss the initial conditions
and parameters. We also explain the evolution of the scalar fields
and the order of events that lead to the final baryon asymmetry. A
quantitative treatment of the scenario can be found in Sec.~\ref{sec:Analytic-Estimate},
where we give analytical estimates for the final baryon asymmetry
of the scenario while we also present numerical results for a large
range of initial conditions. In Sec.~\ref{sec:fermions} we discuss
several possibilities for the needed coupling between the SM and the
newly introduced scalar sector. In Sec.~\ref{sec:Stability-QBalls}
we study the stability of our model concerning particle processes
and inhomogeneities in the scalar condensates, that might lead to
Q-Ball formation. Finally, we present our conclusions in Sec.~\ref{sec:Conclusions}.

\section{\label{sec:Scenario}The Scenario}

Our scenario begins at the end of inflation, which ensures a flat
and homogeneous Friedmann-Robertson-Walker space-time. Ignoring SM
particles and right-handed neutrinos for the moment the Lagrangian
$\mathcal{L}$ of the system is given by \begin{eqnarray}
\mathcal{L} & = & \kl{\frac{1}{2}}(\partial_{\mu}\phi)^{*}(\partial^{\mu}\phi)-V(|\phi|)\nonumber \\
 & + & \kl{\frac{1}{2}}(\partial_{\mu}\chi)^{*}(\partial^{\mu}\chi)-\kl{\frac{1}{2}}\mu_{\chi}^{2}|\chi|^{2}\label{lag}\\
 & - & \kl{\frac{1}{2}}\lambda_{1}|\phi|^{2}|\chi|^{2}-\kl{\frac{1}{4}}\lambda_{2}(\phi^{2}\chi^{*2}+\mbox{h.c.}),\nonumber \end{eqnarray}
 with the quintessence field $\phi$ and a mediating field $\chi$.
For stability reasons we require $\lambda_{1}\geq\lambda_{2}\geq0$.
While the $\chi$ field will later on be coupled to the fermionic
sector, which results in an additional term in its equation of motion,
we take the behavior of the quintessence field to be completely determined
by the classical equations of motion derived from this Lagrangian.
This is equivalent to the assumption that all interactions of~$\phi$
are already included in its effective potential $V(\phi)$ and the
given couplings to the $\chi$ field. This will be further motivated
in section \ref{sec:Stability-QBalls}. 

One can see that the system has a global $U(1)$-symmetry. Later on,
we will choose the fermionic couplings in such a way that we can identify
the corresponding conserved quantity with lepton number~$L$ and
assign both fields the charge~$L=-2$. 
Therefore, the postulated $U(1)$-symmetry for the complete model is just \BmL{},
which is already inherent in the SM.
The lepton number
density stored in the quintessence condensate is 
in this case 
given by \begin{equation}
n_{\phi}=-i(\dot{\phi}^{*}\phi-\phi^{*}\dot{\phi})=-2|\phi|^{2}\dot{\theta},\end{equation}
 where $\theta$ is the phase of $\phi=|\phi|e^{i\theta}$. Then the
lepton number density per comoving volume $A_{\phi}$ has the form\begin{equation}
A_{\phi}=-2|\phi|^{2}\dot{\theta}\left(\frac{a(t)}{a_{0}}\right)^{\!3}\!\!\!,\label{Aphidef}\end{equation}
 where~$a(t)$ is the cosmic scale factor and $a_{0}\equiv a(t_{0})$.
Furthermore, the quantities~$n_{\chi}$ and~$A_{\chi}$ corresponding
to the mediating field~$\chi=|\chi|e^{i\sigma}$ with the phase function~$\sigma$
are defined analogously. 

For the quintessence potential we choose a negative exponential form~\cite{Barreiro:1999zs,Ferreira:1997au},
\begin{equation}
V(|\phi|)=V_{0}\left(e^{-\xi_{1}|\phi|/M_{\textrm{Pl}}}+ke^{-\xi_{2}|\phi|/M_{\textrm{Pl}}}\right)\!\!,\label{pot}\end{equation}
 where $M_{\textrm{Pl}}\equiv1/\sqrt{8\pi G}$ is the reduced Planck
mass. 
This is a typical quintessence potential which leads to
a light quintessence mass and extremely small higher derivatives
at late times.
Also, for typical parameter values ($\xi_{1}=\mathcal{O}(10)$, $\xi_{2}=\mathcal{O}(1)$,
$k\ll1$) such a potential yields a tracker behavior, where the energy
density of the quintessence field follows the background energy density.
The dark energy and cold dark matter densities are
directly related by a constant ratio of order $3/\xi_1^2$ \cite{Ferreira:1997au}
during matter domination in these models. 
The second term in the potential yields the transition to the late-time
acceleration. Since our work concerns only early times, we set~$k=0$
for the rest of this paper. Also many other models lead to equivalent
potentials at early times, see e.g.\  Refs.~\cite{Albrecht:1999rm,Hebecker:2000zb}.
Typical values of the quintessence vacuum expectation value at late
times are of the order of the Planck scale and beyond. 

Our scenario starts at the end of inflation with an oscillating inflaton
and Hubble rate~$H_{\textrm{Inf}}$. The amplitude of the inflaton
oscillations will first decay due to expansion of the universe and
later on due to the decay of the inflaton condensate during reheating
($H=H_{\textrm{R}}$). Obviously these conditions are fulfilled by
most inflationary scenarios and a dependence of our scenario on the
particular inflationary model only comes into play when quantitatively
calculating the produced amount of the final baryon asymmetry. 

At the beginning $(t=t_{0}\equiv0)$ both fields have initial VEVs
of respective absolute amounts $\phi_{0}$,~$\chi_{0}$ and a relative
phase~$\alpha_{0}$. Similar initial conditions were considered in
Refs.~\cite{Balaji:2004xy,Balaji:2005ha} (see also Refs.~\cite{Dolgov:1991fr,Malquarti:2002bh}).
We assume the VEVs to be of such kind that the main part of the energy
density is still given by the inflaton. 

The equations of motion resulting from the Lagrangian~(\ref{lag})
are given by\begin{eqnarray}
\ddot{\phi}+3H\dot{\phi} & = & -2\frac{\textrm{d}V}{\textrm{d}\phi^{*}}-\lambda_{1}|\chi|^{2}\phi-\lambda_{2}\phi^{*}\chi^{2},\label{eomphi}\\
\ddot{\chi}+3H\dot{\chi}+\Gamma\dot{\chi} & = & -(\mu^{2}+\lambda_{1}|\phi|^{2})\chi-\lambda_{2}\chi^{*}\phi^{2},\label{eomchi}\end{eqnarray}
 where we added an additional damping factor~\mbox{$\Gamma\equiv g^{2}m_{\chi}/(8\pi)$}
due to the fermionic couplings of~$\chi$ \cite{Linde:2005ht,Kofman:1997yn},
with~$g^{2}$ being the sum of the squares of the coupling constants
of these interactions and \mbox{$m_{\chi}\equiv(\mu_{\chi}^{2}+\lambda_{1}|\phi|^{2})^{1/2}$}
being the effective mass of~$\chi$.%
\footnote{Even though different directions of $\chi$ might have different masses,
the definition of $\Gamma$ is perfectly fine, as long as we are only
concerned with orders of magnitude and $\lambda_{1}$ and $\lambda_{2}$
are not fine-tuned (see also Sec.~\ref{sec:Analytic-Estimate}).
For the same reason possible decays of $\chi$ into $\phi$ can also
be neglected (see Sec.~\ref{sec:Stability-QBalls}).%
} In order for the interaction of $\phi$ and $\chi$ to have a noticeable
effect we require $\mu_{\chi}^{2}\ll\lambda_{1}|\phi|^{2}$ at all
times. We also stress, that whenever this damping term comes into
effect the relation $|\dot{\phi}/\phi|\ll\phi$ is fulfilled and we
can consider $m_{\chi}$ as approximately constant. 

Since the last term in brackets of Eq.~(\ref{lag}) can be written
as $\frac{1}{2}\lambda_{2}|\phi|^{2}|\chi|^{2}\cos(2\alpha)$, with
the phase difference \mbox{$\alpha\equiv\theta-\sigma$}, we see
that both fields will develop non-zero phase-velocities if the initial
conditions fulfill \mbox{$\sin(2\alpha_{0})\ne0$}, corresponding
to a spontaneous $CP$-violation, see also Ref.~\cite{Balaji:2004xy}. 

In the case of negligible fermionic couplings ($\Gamma=0$) lepton
number is conserved in the scalar field sector and as long as the
homogeneous condensates do not decay the phase space rotation of~$\phi$
and~$\chi$ will always fulfill the constraint \mbox{$A_{\phi}=-A_{\chi}$}. 

For sizeable fermionic couplings of~$\chi$ this simple relation
no longer remains valid. Using Eq.~(\ref{Aphidef}) and the equations
of motion~(\ref{eomphi}), and~(\ref{eomchi}) we now find \begin{equation}
\frac{\textrm{d}}{\textrm{d}t}(A_{\phi}+A_{\chi})=-\Gamma A_{\chi},\label{asym}\end{equation}
 which shows that the explicit development of the fields has to be
considered, which in return depends on their initial conditions at
the end of inflation. For the moment we will fix them in a way where
one can easily see how a baryon-asymmetry in the fermion sector can
arise. However, the system is unlikely to be restricted to these initial
values and we will consider a much broader parameter range in the
following section. With this in mind, we put $\lambda_{2}$ of order
one tenth, all other coupling constants of order one and both initial
field values of the order of $H_{\textrm{Inf}}$ at the end of inflation.
Additionally, we choose $V_{0}$ in such a way that \mbox{$H_{\textrm{Inf}}^{2}M_{\textrm{Pl}}^{2}\gg V(\phi\approx H_{\textrm{Inf}})\geq H_{\textrm{Inf}}^{3}M_{\textrm{Pl}}$},
which ensures that the main part of the background energy density
at the beginning of the scenario ($\rho_{0}^{\textrm{crit}}\equiv3H_{\textrm{Inf}}^{2}M_{\textrm{Pl}}^{2}$)
is given by the inflaton, while the behavior of $\phi$ is still governed
by $V(\phi)$ and not by its mixing terms with $\chi$. 

With the above mentioned, $\phi$ will now rapidly increase right
from the start for typical parameter values in the quintessence potential.
Since the mixing terms are of the order of the Hubble scale, the phase
velocities of both fields also become sizeable. When $|\phi|$ grows
big enough such that $\Gamma\approx H$ the $\chi$-condensate starts
to decay due to its fermionic couplings and its oscillation amplitudes
rapidly decrease. Since these interactions are lepton number conserving
the decay also mediates a lepton-number transfer to the fermionic
sector. The conclusive size of the transferred lepton asymmetry $A_{\infty}$
can easily be found from the fact that total lepton number per comoving
volume is conserved and that the $\chi$-condensate has decayed at
late times: \begin{equation}
A_{\infty}\equiv-A_{\phi}|_{t\rightarrow\infty}=\int_{0}^{\infty}\textrm{d}t\cdot\Gamma A_{\chi},\label{eq:asymdecay}\end{equation}
 where we additionally used equation~(\ref{asym}). 

Finally $B$-~and~$L$-violating electroweak sphaleron processes
transform parts of the lepton asymmetry in the fermionic sector, such
that the baryon asymmetry is of the same order of magnitude as the
lepton asymmetry in the DE sector~\cite{Harvey:1990qw}.

\section{\label{sec:Analytic-Estimate}Quantitative Treatment}

In order to show that the final asymmetry, which is produced in our
scenario, can quantitatively explain the observed value of the baryon
to entropy ratio in the universe, it is desirable to get an analytic
estimate of the asymptotic value $A_{\infty}=-A_{\phi}|_{t\rightarrow\infty}$,
which we will do in the first part of this section. We will then use
this estimate to quantitatively determine the final baryon asymmetry
of the scenario while we will also present numerical results. 

An analytic estimate is indeed possible for a wide range of parameters
and initial conditions by solving the equation of motion (\ref{eomchi})
of the $\chi$-field approximately using a WKB approach, and then
calculating $A_{\infty}$ through the integral in Eq.~(\ref{eq:asymdecay}).
Here, we restrict ourselves to the parameter space for which the relation \begin{equation}
|\theta(t\rightarrow\infty)-\theta(t=0)|\leq\int_{0}^{\infty}\textrm{d}t\,\frac{|A_{\phi}|}{2(a/a_{0})^{3}|\phi|^{2}}\ll1\label{wkbphaseapprox}\end{equation}
is valid. This will be true for the biggest part of the considered values due to the growing VEV of the quintessence field. In this case, we can approximate $\theta$ to be constant and therefore take the
quintessence field to be along the real axis without loss of generality.%
\footnote{Let us emphasize that this a very good approximation for the dynamics
of the $\chi$-field, but does not mean that the phase velocity could
be neglected as far as the dynamics of the whole system are concerned,
especially when calculating $A_{\phi}$ via Eq.~(\ref{Aphidef})
numerically. The exact parameter space where Eq.~(\ref{wkbphaseapprox}) and hence the WKB approximation
is valid is discussed later in this section, where we also consider further parameter ranges numerically.%
} 
Now, the equation of motion (\ref{eomchi}) of the $\chi$-field
can be split into two independent linear equations for the real and
imaginary parts $\chi_{1}$ and $\chi_{2}$ which are of the form of
a harmonic oscillator with time-dependent damping \mbox{$\gamma(t)\equiv3H(t)+\Gamma(t)$}
and frequencies \mbox{$\omega_{1,2}^{2}(t)\equiv(\lambda_{1}\pm\lambda_{2})\phi^{2}(t)+\mu_{\chi}^{2}$},
respectively. For simplicity, we will restrict ourselves to the case
where $\gamma\ll\omega_{1,2}$ keeping the oscillator from being overdamped
and assume that the VEV of the quintessence field gives the main contribution
to the mass of the $\chi$ field, $\mu_{\chi}^{2}\ll(\lambda_{1}\pm\lambda_{2})\phi_{0}^{2}$.
The requirement for the validity of the WKB approximation is that
the rate of change of the frequency and the damping term are much
smaller than the frequency itself, which in our case simply means
\begin{equation}
|\dot{\phi}/\phi|\ll|\phi|.\label{wkbabsapprox}\end{equation}

\begin{figure}
\begin{center}\includegraphics[%
  clip,
  width=1\columnwidth,
  keepaspectratio]{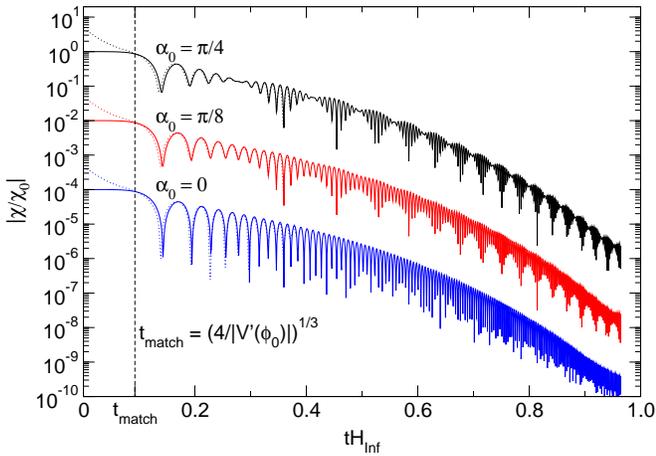}\vspace{-0.5cm}\end{center}

\caption{\label{Chifield}\textit{\small Absolute value $|\chi(t)|$ (solid),
and the WKB approximation (dotted) for $H_{\textrm{Inf}}=\phi_{0}=\chi_{0}=10^{12}\,\textrm{GeV}$,
$g=1$, $\lambda_{1}=1$, $\lambda_{2}=0.1$, $V_{0}=10^{-4}\rho_{0}^{crit}$,
$\xi_{1}=7$ and different values of $\alpha_{0}$. For reasons of
clearness the plot has been rescaled by factors $10^{-2}$ and $10^{-4}$
for $\alpha_{0}=\pi/8$ and $0$, respectively. For $\alpha_{0}\not=0$
there is a beat frequency $\omega_{-}/2\equiv(\omega_{1}-\omega_{2})/2$
corresponding to the large oscillations in~$|\chi|$, while the fast
oscillations of $\chi$ itself are set by $\omega_{+}/2\equiv(\omega_{1}+\omega_{2})/2$.}}
\end{figure}

For the moment we will assume this equation to hold, while the validity of this assumption will be discussed later in this section.
Now, the result of the WKB approximation to leading order in $\dot{\phi}/\phi^{2}$
and $\gamma/\omega_{1,2}$ is given by (see also Fig.~\ref{Chifield})\begin{equation}
\frac{\chi_{1,2}(t)}{\chi_{1,2}^{0}}=\sqrt{\frac{\omega_{1,2}(0)}{\omega_{1,2}(t)}}\, e^{{\displaystyle \!\kl{-\frac{1}{2}\int_{0}^{t}\!\textrm{d}\tilde{t}\,\gamma(\tilde{t})}}}\cos\left(\int_{0}^{t}\!\!\textrm{d}\tilde{t}\,\omega_{1,2}(\tilde{t})\right)\!\!,\label{wkbchi}\end{equation}
 where the initial condition $\dot{\chi}(0)=0$ has been imposed and
$\chi(0)=\chi_{0}e^{-i\alpha_{0}}$ is satisfied by the choices $\chi_{1}^{0}=\chi_{0}\cos(\alpha_{0})$
and $\chi_{2}^{0}=-\chi_{0}\sin(\alpha_{0})$.%

\begin{figure}
\begin{center}\includegraphics[%
  clip,
  width=1\columnwidth,
  keepaspectratio]{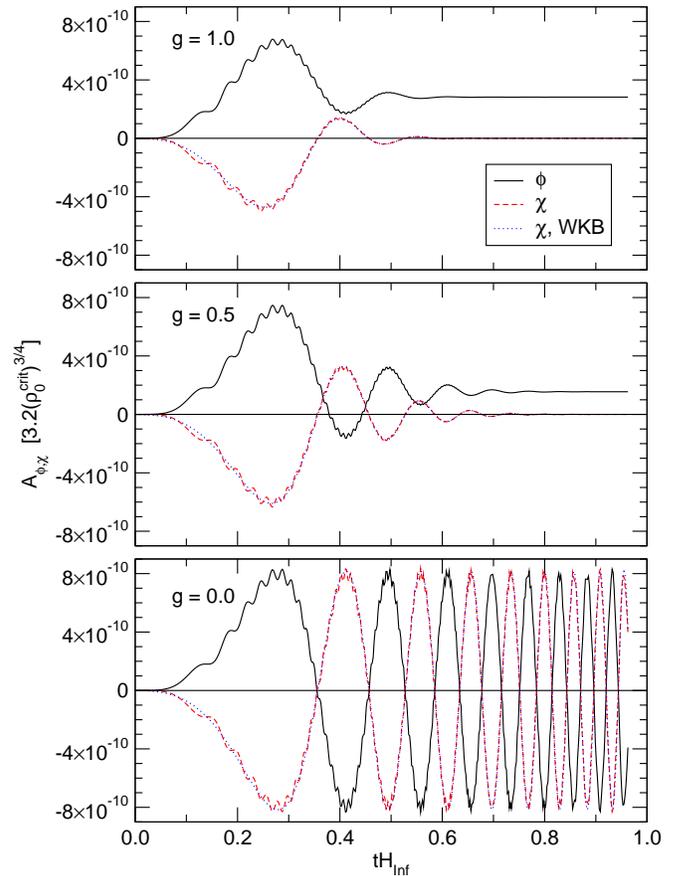}\vspace{-0.5cm}\end{center}

\caption{\label{Asymmetry}\textit{\small The comoving asymmetries $A_{\phi,\chi}=n_{\phi,\chi}(a/a_{0})^{3}$
in units of $[3.2\,(\rho_{0}^{\textrm{crit}})^{3/4}]$ for $\phi$
(solid), $\chi$ (dashed) and the WKB approximation (dotted, only
the leading term proportional to $\omega_{+}$ in the large bracket
in Eq.~(\ref{wkbAchi}) has been kept) for $\alpha_{0}=-\pi/4$ and
the other parameters as in Fig.~\ref{Chifield}. In the two upper
graphs the $\chi$ field decays ($g=1,\,0.5$), the lower graph shows
the behavior without decay for comparison. In these units, the \BmL{}
asymmetry $n/s$ is the asymptotic value of $A_{\phi}$ for $D=1$
according to Eq.~(\ref{ratioentropy}).}}
\end{figure}

Thus $\chi_{1,2}(t)$ are both oscillating, but with different frequencies
$\omega_{1,2}$ if $\lambda_{2}\not=0$. This leads to a modulation
in the phase velocity of the complex field $\chi$ and thus of the
comoving asymmetry, which can be calculated using the corresponding
expression to Eq.~(\ref{Aphidef}) for $\chi$:\begin{equation}
\begin{array}{ll}
{\displaystyle A_{\chi}(t)} & {\displaystyle \!\!=\frac{1}{2}\sin(2\alpha_{0})\,\chi_{0}^{2}\,\sqrt{\frac{\omega_{1}(0)\omega_{2}(0)}{\omega_{1}(t)\omega_{2}(t)}}\, e^{\kl{-\!\int_{0}^{t}\textrm{d}\tilde{t}\,\Gamma(\tilde{t})}}}\\
 & \!\!\times{\displaystyle \left(\omega_{+}\sin\left(\int_{0}^{t}\!\!\textrm{d}\tilde{t}\,\omega_{-}(\tilde{t})\right)-\omega_{-}\sin\left(\int_{0}^{t}\!\!\textrm{d}\tilde{t}\,\omega_{+}(\tilde{t})\right)\right)}\end{array}\label{wkbAchi}\end{equation}
 with $\omega_{\pm}\equiv\omega_{1}\pm\omega_{2}=(\sqrt{\lambda_{1}+\lambda_{2}}\pm\sqrt{\lambda_{1}-\lambda_{2}})\phi(t)$.
Note that only the second part of the damping term%
\footnote{It is now easy to convince oneself that even if different decay widths
for $\chi$ in different directions in the complex plane were assumed
the only change in the resulting asymmetry $A_{\chi}$ would be to
replace $\Gamma$ in Eq.~(\ref{wkbAchi}) by the mean value of the
corresponding decay widths, without changing the main result.%
} $\gamma(t)=3H(t)+\Gamma(t)$ in Eq.~(\ref{wkbchi}), which comes
from the decay of $\chi$, appears in the asymmetry, while the Hubble
term gives a contribution $(a_{0}/a(t))^{3/2}$ which is exactly canceled.
Thus, if there was no decay, $\Gamma=0$, the amplitude of the oscillations
in the (comoving) asymmetry would stay constant, whereas the frequency
increases since the VEV of the quintessence field $\phi$ grows when
it rolls down its potential, see the lower graph in Fig.~\ref{Asymmetry}.
When $\Gamma>0$, the increase in the frequency still occurs, but
the amplitude is damped by the exponential term in Eq.~(\ref{wkbAchi}),
see the two upper graphs in Fig.~\ref{Asymmetry}. Furthermore, the
asymmetry is proportional to $\sin(2\alpha_{0})$, i.e.\  it is zero
if there is no initial phase difference ($\alpha_{0}=0$), and its
modulus is maximal if $\alpha_{0}=\pm\pi/4$.%

It is instructive to consider the case when $\lambda_{2}\ll\lambda_{1}$,
because then $\omega_{+}^{2}\approx4\lambda_{1}\phi^{2}\gg\omega_{-}^{2}\approx(\lambda_{2}^{2}/\lambda_{1})\phi^{2}$
and the first term in the large brackets of Eq.~(\ref{wkbAchi})
dominates, giving the leading oscillations of $A_{\chi}$ with a {}``low''
frequency $\omega_{-}$ proportional to the parameter $\lambda_{2}$,
superimposed by a {}``fast'' oscillating component whose amplitude
is suppressed by a relative factor $\omega_{-}/\omega_{+}$ (see Fig.~\ref{Asymmetry}). 

Finally, the analytic expression (\ref{wkbAchi}) together with a
decay rate of the form $\Gamma(t)=g^{2}/(8\pi)\sqrt{\lambda_{1}}\phi(t)$
can be used to calculate the total asymptotic asymmetry (per comoving
volume)~$A_{\infty}$ via the integral in Eq.~(\ref{eq:asymdecay}): 

\begin{equation}
A_{\infty}=\frac{1}{2}\sin(2\alpha_{0})\,\chi_{0}^{2}\,\phi_{0}\sqrt{\lambda_{1}}\frac{g^{2}}{8\pi}\frac{\omega_{+}\omega_{-}}{\Gamma^{2}+\omega_{-}^{2}}\left(1-\frac{\omega_{-}^{2}}{\omega_{+}^{2}}\right)\!\!.\label{finalasym}\end{equation}
 Note that the time-dependence of $\phi(t)$ disappears in the final
result and only the initial values $\phi_{0}$ and $\chi_{0}$ enter
as dimensionfull quantities. Furthermore, $A_{\infty}$ vanishes if
either $\Gamma\propto g^{2}/(8\pi)\rightarrow0$ or $\lambda_{2}\propto\omega_{+}\omega_{-}\rightarrow0$
since in these limiting cases no transfer of asymmetry between $\chi$
and the SM, or, respectively, $\chi$ and the quintessence sector
would be possible. If, on the other hand, $g^{2}/(8\pi)\approx\lambda_{2}/\lambda_{1}\ll1$,
both transfers are of comparable {}``strength'' and the asymmetry
is simply $A_{\infty}\approx\sin(2\alpha_{0})\,\chi_{0}^{2}\,\sqrt{\lambda_{1}}\phi_{0}/2$,
independent of $g$ and $\lambda_{2}$. For given initial values $\phi_{0},\chi_{0},\alpha_{0}$
and fixed $\lambda_{1}$, this is the maximum asymmetry that can be
transferred. 

\begin{figure}
\begin{center}\vspace{-0.9cm}\includegraphics[%
  clip,
  width=1\columnwidth,
  keepaspectratio]{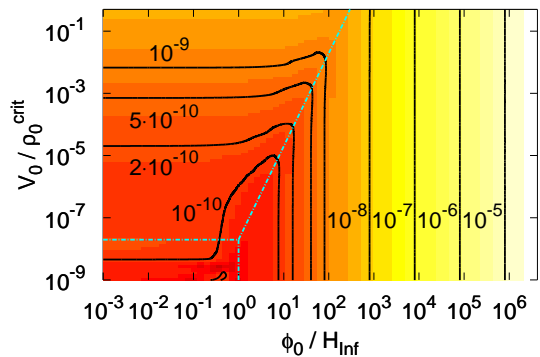}\vspace{-0.9cm}\end{center}

\caption{\label{Contourplot}\textit{\small This contour plot shows the final
asymmetry factor \mbox{$\kappa$} from Eq.~(\ref{ratioentropy})
for the same parameters as in Fig.~\ref{Asymmetry} ($g=1$) apart
from varying values of the initial quintessence VEV $\phi_{0}$ and
the initial potential energy $V_{0}$. The light dot-dashed lines
divide the plot into three domains: In the right part, due to Eq.~(\ref{WKBwomatching}),
$\phi_{0}^{3}\gg\chi_{0}^{2}\phi_{0},\,\,|V'(\phi_{0})|$, the asymmetry
depends on $\phi_{0}$ but not on $V_{0}$, whereas well inside the
upper left part it is vice versa since $|V'(\phi_{0})|\gg\phi_{0}^{3},\,\,\chi_{0}^{3}$
from Eq.~(\ref{WKBmatching}). This is in accordance with the WKB
results~(\ref{wkbresults}) inside each of these domains. In the
remaining part in the lower left corner backreaction effects of $\chi$
on $\phi$ lead to more complex dynamics. Note that $n/s=\kappa D$
with $D=1\dots10^{-6}$, see Eqs.~(\ref{ratioentropy}) and~(\ref{kapparange}).}}
\end{figure}

It turns out that Eq.~(\ref{finalasym}) gives a very accurate estimate
for~$A_{\infty}$ in the case \begin{equation}
\phi_{0}^{3}\gg\chi_{0}^{2}\phi_{0},\,\,|V'(\phi_{0})|,\label{WKBwomatching}\end{equation}
 since conditions (\ref{wkbphaseapprox}) and (\ref{wkbabsapprox})
are both fulfilled from the beginning ($t=0$). 
This in return can easily be seen from the approximate analytic solution for the quintessence
field for early times (when $\phi(t)\ll\Mplr$) \begin{equation}
\phi(t)\approx\phi_{0}+\frac{1}{2}|V'(\phi_{0})|t^{2}.\label{earlyphi}\end{equation}
In this domain the
asymmetry is independent of $V_{0}$, but (for fixed $\chi_{0}$)
only depends on the initial quintessence VEV $\phi_{0}$ (corresponding
to the vertical contour lines in Fig.~\ref{Contourplot}).%

If, on the other hand, \begin{equation}
|V'(\phi_{0})|\gg\phi_{0}^{3},\,\,\chi_{0}^{3},\label{WKBmatching}\end{equation}
condition~(\ref{wkbphaseapprox}) on the phase is
actually also satisfied, but condition~(\ref{wkbabsapprox})
is only fulfilled for times $t\gtrsim t_{\textrm{match}}\equiv(4/|V'(\phi_{0})|)^{1/3}$
as can again be easily seen using (\ref{earlyphi}).
To get an estimate of the asymmetry in this case one can use the early-time
solution (\ref{earlyphi}) for $t<t_{\textrm{match}}$ and then match
the WKB solution for $t>t_{\textrm{match}}$. Since $\phi_{0}t_{\textrm{match}}\ll1$
(see Eq.~(\ref{WKBmatching})) the $\chi$-field can be approximated
to stay static until $t_{\textrm{match}}$, and thus the only modification
in the WKB results obtained above, especially in the final asymmetry
(\ref{finalasym}), is to replace the initial time $t=0$ by $t=t_{\textrm{match}}$
and \begin{equation}
\phi_{0}\rightarrow\phi(t_{\textrm{match}})\approx\sqrt[3]{2|V'(\phi_{0})|}=\sqrt[3]{2\xi_{1}V_{0}/\Mplr}.\label{phimatch}\end{equation}
 In this domain, in contrast to the previous case, the final asymmetry
is thus independent of the initial quintessence VEV $\phi_{0}$ and
only depends on the initial potential energy%
\footnote{For $\phi_{0}\ll\Mplr$ we have $V(\phi_{0})=V_{0}$.%
} $V_{0}$ (corresponding to the horizontal contour lines in Fig.~\ref{Contourplot}). 

To determine the final baryon number, we need to calculate the ratio
of the produced \BmL{} asymmetry $n\equiv-A_{\infty}(a_{0}/a)^{3}$
and the entropy density~\cite{Kolb:1990vq},\[
s=\frac{2\pi^{2}}{45}g^{*}(T_{\textrm{R}})T_{\textrm{R}}^{3}\left(\frac{a_{\textrm{R}}}{a}\right)^{3}\approx3.2\,\rho_{\textrm{R}}^{3/4}\left(\frac{a_{\textrm{R}}}{a}\right)^{3}\!\!\!\!,\]
 where $T_{\textrm{R}}$, $\rho_{\textrm{R}}=g^{*}(T_{\textrm{R}})\pi^{2}T_{\textrm{R}}^{4}/30$
and $a_{\textrm{R}}$ are the temperature, energy density and the
scale factor at reheating, respectively, and the number $g^{*}(T_{\textrm{R}})$
of active degrees of freedom has been assumed to be of order one hundred.
In principle, the final asymmetry of the scenario is independent of
the physics between inflation and reheating, however, inflationary
dynamics can enter indirectly through the expansion rate when relating
the corresponding scale factors and critical energy densities via
\[
\rho_{\textrm{R}}^{3/4}a_{\textrm{R}}^{3}=(\rho_{0}^{\textrm{crit}})^{3/4}a_{0}^{3}\exp[\kl{\frac{3}{4}\int_{0}^{t_{\textrm{R}}}\textrm{d}t\, H(t)(1-3\omega(t))}],\]
 where $\omega(t)$ is the equation of state parameter of the dominant
component, i.e.\  the inflaton and later radiation. Thus the final
asymmetry relative to the entropy density is given by%
\footnote{Since only a part of this asymmetry is transferred from the leptonic
to the baryonic sector there will be an additional sphaleron factor
of order one, see Ref.~\cite{Harvey:1990qw}.%
}\begin{equation}
\frac{n}{s}=\underbrace{\frac{-A_{\infty}}{3.2(\rho_{0}^{\textrm{crit}})^{3/4}}}_{\equiv\,\kappa}\,\underbrace{\exp\left({\displaystyle -\frac{3}{4}\int_{0}^{t_{\textrm{R}}}\!\textrm{d}t(1-3\omega(t))H(t)}\right)}_{\equiv\, D}\!\!.\label{ratioentropy}\end{equation}
 The first factor $\kappa$ depends only on the dynamics of the quintessence
and the mediating field shortly after inflation, whereas the second
factor $D$ describes the dilution of the asymmetry between the end
of inflation and the time when $\omega(t)\rightarrow1/3$. In fact,
$1/D$ is equal to the spatial expansion of the universe after inflation
relative to a radiation dominated cosmos. Note that the upper limit
in the integration can be formally extended beyond reheating, since
from then on $\omega=1/3$. This makes it obvious that any inflationary
scenario can only have an impact on the final baryon number through
its equation of state and not through further characteristics, e.g.\ 
the reheating temperature%
\footnote{Of course, there are inflationary scenarios in which the time of reheating
also marks a change in the equation of state, in which case the amount
of the produced baryon asymmetry and the reheating temperature are
not completely independent.%
}. 

To get a plausible range for $D$ we will consider various possibilities:
for example, if the inflaton field $I$ oscillates in a potential
of the form $I^{m}$ with even $m$, its mean equation of state (averaged
over a whole oscillation period) is $\omega=\frac{m-2}{m+2}=\textrm{const}$~\cite{Shtanov:1994ce}.
Assuming that this is true for the whole time between the end of inflation
and reheating we get $D=(2.4\, T_{\textrm{R}}/(\rho_{0}^{\textrm{crit}})^{1/4})^{(4/m-1)}$.
For $T_{\textrm{R}}=10^{9}\,\textrm{GeV}$, $H_{\textrm{Inf}}=10^{12}\,\textrm{GeV}$
and $m=2$ this gives $D\approx10^{-6}$, whereas for $m=4$ there
is no dilution at all yielding $D=1$, with or without preheating~\cite{Shtanov:1994ce}.
Another example is a model with preheating and a quadratic minimum
($m=2$) for the inflaton. For the case recently discussed in Ref.~\cite{Podolsky:2005bw},
one finds $D\approx10^{-1}\dots10^{-3}$. Here we will adopt the point
of view that the dilution factor $D$ will be roughly in the range
$1\dots10^{-6}$, yielding a range for $\kappa$\begin{equation}
10^{-10}\lesssim\kappa\lesssim10^{-4},\label{kapparange}\end{equation}
 that is compatible with the observed value \mbox{$n/s\sim10^{-10}$}~\cite{Spergel:2003cb}.
Using the analytic estimates~(\ref{finalasym}) and~(\ref{phimatch})
one obtains \begin{equation}
\kappa\approx-\frac{\mathcal{N}}{2}\sin(2\alpha_{0})\left(\frac{\chi_{0}}{H_{\textrm{Inf}}}\right)^{\!2}\!\!\times\left\{ \begin{array}{l}
\frac{\phi_{0}}{7.3H_{\textrm{Inf}}}\left(\frac{H_{\textrm{Inf}}}{\Mplr}\right)^{3/2}\\
\frac{1}{4}\left(\frac{\xi_{1}V_{0}}{\rho_{0}^{\textrm{crit}}}\right)^{1/3}\left(\frac{H_{\textrm{Inf}}}{\Mplr}\right)^{7/6}\end{array}\right.\label{wkbresults}\end{equation}
 with $\mathcal{N}\equiv\frac{g^{2}\sqrt{\lambda_{1}}}{8\pi}\frac{\omega_{+}\omega_{-}}{\Gamma^{2}+\omega_{-}^{2}}(1-\omega_{-}^{2}/\omega_{+}^{2})$.
The upper and lower line of Eq.~(\ref{wkbresults}) correspond to
the cases~(\ref{WKBwomatching}) and~(\ref{WKBmatching}), respectively.
For coupling constants $g^{2}/(8\pi)\sim\lambda_{2}/\lambda_{1}\ll1\sim\lambda_{1}$,
$\mathcal{N}$ is of order one. Generically assuming $\sin(2\alpha_{0})\sim-1$
and $\chi_{0}\sim H_{\textrm{Inf}}\sim10^{12}\,\textrm{GeV}\sim4\cdot10^{-7}\Mplr$
we thus find values for $\kappa$ of roughly $10^{-11}\phi_{0}/H_{\textrm{Inf}}$
and $10^{-8}(V_{0}/\rho_{0}^{\textrm{crit}})^{1/3}$ respectively,
which lie inside the allowed range (\ref{kapparange}) for a huge
variety of initial conditions of the quintessence VEV and its potential
energy after inflation (see Fig.~\ref{Contourplot}).

\section{\label{sec:fermions}Fermionic couplings}

There are various possibilities to couple the mediating field~$\chi$
of our toy model to the fermionic sector for which we will give two
examples in this section. 

Since the whole scenario is \BmL{} conserving, the right-handed neutrinos
do not have a Majorana mass term and carry lepton number one. As the
$\chi$-field is also a gauge singlet it can be coupled to one (or
more) right-handed neutrino(s) $N$ via \begin{equation}
g\,\overline{N^{C}}N\chi+\mbox{h.c.}.\end{equation}
 This would assign the desired lepton number to the $\chi$ field
and implies the given decay term~$\Gamma\dot{\chi}$ in the equation
of motion~(\ref{eomchi}). However, if this was the only additional
coupling, the asymmetry would remain stored in the right-handed neutrino
sector due to their tiny Yukawa couplings \cite{Dick:1999je}. In
this case one would have to find a way to transfer the lepton asymmetry
to the SM sector, where sphalerons could create the final baryon asymmetry.
This could be realized by a light scalar field $S$ (with mass $H_{\textrm{Inf}}\gg m_S \gtrsim \rm{TeV} $) with the same
quantum numbers as the SM Higgs field and a quadratic potential. If this scalar would start with a typical VEV of order $H_{\textrm{Inf}}$ at the end of inflation, it would remain frozen in until $H$ drops below the mass of the scalar. In this case an additional coupling term \begin{equation}
g'\,\overline{\ell}NS+\mbox{h.c.}\end{equation}
 with a left-handed lepton doublet $\ell$ and suitable coupling constant
$g'$ should lead to the required left-right-equilibration of the
neutrinos as $H$ becomes smaller.%
\footnote{The quartic potential of the SM Higgs and its small Yukawa couplings to neutrinos in the case of Dirac neutrinos seem to prevent it from being a suitable candidate for this job.} 
When $H$ drops below the mass of the scalar the VEV of $S$ would start oscillating and finally vanish. It might even be possible to avoid the introduction of
such a new particle by introducing a similar coupling between the inflaton
and the neutrinos. In this latter case one would get additional constraints
for the inflationary scenario. 

One possibility, which works without the introduction of additional
particles while also being independent of the inflationary model,
would be to couple the $\chi$ field directly to the left-handed neutrinos
$\nu$ via the term\begin{equation}
g\,\overline{\nu^{C}}\nu\chi+\mbox{h.c.}.\end{equation}
 Due to the heavy mass of $\chi$ this should be consistent with observations.
However, the explicit breaking of gauge invariance makes this option
less attractive.

\section{\label{sec:Stability-QBalls}Stability and Q-Balls}

In this section we discuss possible sources of instability in our
model. For instance, particle processes and spatial inhomogeneities
could disturb the scalar field evolution described in the previous
sections. One peculiarity of complex scalar fields with a conserved
global quantum number is the potential formation of so-called Q-Balls
\cite{Coleman:1985ki}. A Q-Ball is a non-topological soliton and
represents a spatially extended scalar field configuration carrying
a non-zero charge associated to the global symmetry. For our scenario
it is essential that the complex quintessence field does not promote
the formation of Q-Balls, which could make the field unstable and
inhomogeneous and thereby destroy its DE properties. To analyze the
stability of the scalar fields in our model one has to to solve the
corresponding equations of motion including the spatial derivatives
that were omitted in Eqs.~(\ref{eomphi}) and~(\ref{eomchi}). Since
the numerical solution of this system of partial differential equations
is beyond the scope of this paper (see e.g.\ \cite{Kasuya:2001hg}), let us instead proceed with a much
simpler approach that might give us at least some hints towards the
stability properties of our model. Since we also only want to show
that instabilities should not be a problem for our model in general,
we only consider the initial conditions and parameter values as in
Fig. \ref{Chifield} ($g=1$) and neglect perturbations in the metric. 

Following Refs.~\cite{Kusenko:1997si,Kasuya:2001pr} we study the evolution of small
inhomogeneous perturbations in the scalar fields. Substituting $\phi(x,t)\rightarrow\phi(t)+\delta\phi(x,t)$
and $\chi(x,t)\rightarrow\chi(t)+\delta\chi(x,t)$, where the first
parts denote the homogeneous solutions of the previous sections, we
can derive the equations of motion for $\delta\phi$ and $\delta\chi$
from the Lagrangian (\ref{lag}). Keeping only the terms linear in
the small perturbations, we find a mass mixing matrix for $\delta\phi$
and $\delta\chi$, where the dominating contributions on the main
diagonal, at early times, are of the order $V'(|\phi|)/|\phi|$ and
$\lambda_{1}|\phi|^{2}$ for $\delta\phi$ and $\delta\chi$, respectively,
while the off-diagonal terms are of the order $\lambda_{1}\chi\phi$.
As we checked numerically, the off-diagonal entries never become solely
dominant, therefore we will consider the perturbations of $\phi$
and $\chi$ separately in this simplified analysis. 

Now, we write the complex scalars~$\phi$ and~$\chi$ in the form~$\phi=\varphi e^{i\theta}$,~$\chi=|\chi|e^{i\sigma}$
with real functions~$\varphi$,~$\theta$ and~$\sigma$. Within
the quintessence field~$\phi$ we consider perturbations of the type~$\delta\varphi=\delta\varphi_{0}\exp(\Omega(t)+i\vec{k}\vec{x})$,~$\delta\theta=\delta\theta_{0}\exp(\Omega(t)+i\vec{k}\vec{x})$,
that are characterized by a wavenumber~$k=|\vec{k}|$ and a time-dependent
function~$\Omega(t)$. From the Lagrangian~(\ref{lag}) we derive
the equations of motion for the quintessence field \begin{eqnarray}
0 & = & \ddot{\varphi}+3H\dot{\varphi}-a^{-2}\nabla^{2}\varphi-\varphi\dot{\theta}^{2}+a^{-2}\varphi(\nabla\theta)^{2}\nonumber \\
 & + & V'(\varphi)+|\chi|^{2}\varphi[\lambda_{1}+\lambda_{2}\cos2\alpha],\\
0 & = & \ddot{\theta}+3H\dot{\theta}-a^{-2}\nabla^{2}\theta+2\dot{\theta}\kl{\frac{\dot{\varphi}}{\varphi}}\nonumber \\
 & - & 2a^{-2}(\nabla\varphi)(\nabla\theta)-\lambda_{2}|\chi|^{2}\sin2\alpha,\end{eqnarray}
 where~$\alpha=\theta-\sigma$ is the phase difference of both scalar
fields. By substituting~$\varphi\rightarrow\varphi+\delta\varphi$,
$\theta\rightarrow\theta+\delta\theta$ and keeping only terms at
most linear in the perturbations~$\delta\varphi$,~$\delta\theta$,
one obtains a system of equations for~$\delta\varphi$ and~$\delta\theta$.
The condition for this system to have a non-trivial solution reads\begin{eqnarray}
0 & = & \left(\ddot{\Omega}+\dot{\Omega}^{2}+3H\dot{\Omega}+\kl{\frac{k^{2}}{a^{2}}}-\dot{\theta}^{2}+V''(\varphi)\right.\nonumber \\
 & + & |\chi|^{2}[\lambda_{1}+\lambda_{2}\cos2\alpha]\Big)\times\left(\ddot{\Omega}+\dot{\Omega}^{2}\right.\nonumber \\
 & + & \left.3H\dot{\Omega}+\kl{\frac{k^{2}}{a^{2}}}+2\dot{\Omega}\kl{\frac{\dot{\varphi}}{\varphi}}-2\lambda_{2}|\chi|^{2}\cos2\alpha\right)\nonumber \\
 & + & 4\dot{\theta}\left(\dot{\Omega}\dot{\theta}+\lambda_{2}|\chi|^{2}\sin2\alpha\right)\times\left(\dot{\Omega}-\kl{\frac{\dot{\varphi}}{\varphi}}\right)\!.\label{eq:QB-om-full-equ}\end{eqnarray}
 If this equation for~$\Omega(t)$ has any solution with \mbox{$\textrm{Re}(\dot{\Omega})>0$}
for a period of time, then the perturbation mode \mbox{$\sim\exp(\Omega(t)+i\vec{k}\vec{x})$}
is growing exponentially, possibly indicating an instability in the
scalar field~$\phi$, which could lead to the formation of Q-Balls.
The following discussion is divided into three parts. First, during
the early moments of evolution, when the mediating field~$\chi$
has not yet decayed, we will study the stability of the model mainly
numerically. The second and third parts concern the time directly
after the $\chi$-decay and the very late cosmological era, respectively,
where we can now treat the problem completely analytically. 

\begin{figure}
\begin{center}\includegraphics[%
  clip,
  width=1\columnwidth,
  keepaspectratio]{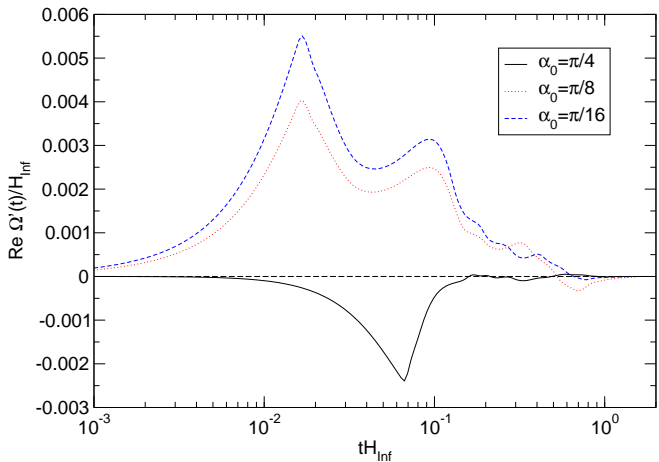}\vspace{-0.5cm}\end{center}

\caption{\label{cap:Pert-Freq} \textit{The real part of~$\dot{\Omega}(t)$,
where the function~$\Omega(t)$ describes the evolution of the inhomogeneous
perturbations $\delta\varphi,\delta\theta\propto\exp(\Omega(t)+i\vec{k}\vec{x})$
for different values of the initial phase difference:~ $\alpha_{0}=\pi/4$,
$\pi/8$, $\pi/16$. For this numerical solution the wavenumber~$(k/a)$
and the initial value of~$\dot{\Omega}$ are taken to be~$0$. The
remaining parameters are the same as in Fig.~\ref{Chifield}.}}
\end{figure}

Starting at the end of inflation we solve Eq.~(\ref{eq:QB-om-full-equ})
numerically for the function~$\dot{\Omega}(t)$ and several different
choices of the wavenumber~$(k/a)^{2}$ and the initial phase difference~$\alpha_{0}$.
In some cases one can observe positive values of~$\textrm{Re}(\dot{\Omega})$,
which appear maximal for small~$k$. As an example the~$k=0$ case
is shown in Fig.~\ref{cap:Pert-Freq}. However, for these unstable
modes,~$\textrm{Re}(\dot{\Omega})$ turns out to be much smaller
than the Hubble scale~$H_{\textrm{Inf}}$ so that there is not enough
time for the instabilities to develop. In addition, a wavenumber smaller
than~$H_{\textrm{Inf}}$ implies that the wavelength of the perturbation
is larger than the Hubble radius and its evolution is therefore suppressed
by the cosmological expansion. For modes with large wavenumbers~$(k/a)^{2}\gg H_{\textrm{Inf}}^{2}$
the numerical solutions of Eq.~(\ref{eq:QB-om-full-equ}) do not show growing
perturbations, as expected. %

To study the late-time behavior of the inhomogeneities, we simplify
Eq.~(\ref{eq:QB-om-full-equ}) by neglecting $|\ddot{\Omega}|\ll|\dot{\Omega}^{2}|$,
i.e.\ assuming a slowly changing $\omega\equiv\dot{\Omega}$ since
the background solution is now slowly varying. In addition the mediating
field~$\chi$ has already decayed~($|\chi|=0$). Thus, we obtain\begin{eqnarray}
0 & = & \left(\omega^{2}+3H\omega+\kl{\frac{k^{2}}{a^{2}}}-\dot{\theta}^{2}+V''(\varphi)\right)\nonumber \\
 & \times & \left(\omega^{2}+3H\omega+\kl{\frac{k^{2}}{a^{2}}}+2\omega\kl{\frac{\dot{\varphi}}{\varphi}}\right)+4\omega\dot{\theta}^{2}\left(\omega-\kl{\frac{\dot{\varphi}}{\varphi}}\right)\!\!,\label{eq:QB-om-simple-equ}\end{eqnarray}
 where solutions with a positive real part of~$\omega$ might indicate
an instability. We find that the range of values for the wavenumber~$k/a$
that solves this equation for~$\textrm{Re}(\omega)>0$ is given approximately
by~\begin{equation}
0<\left(\frac{k}{a}\right)^{\!2}\lesssim\left(\frac{k_{\textrm{max}}}{a}\right)^{\!2}\!\equiv\dot{\theta}^{2}-V''(\varphi).\end{equation}
 Hence, the late-time evolution of the quintessence field is stable
for~$\dot{\theta}^{2}\lesssim V''(\varphi)$. To verify this condition
let us assume for the comoving asymmetry of the quintessence field
a maximal value of~\mbox{$|A_{\phi}|=2|\dot{\theta}|\varphi^{2}(a/a_{0})^{3}\sim10^{-8}(\rho_{0}^{\textrm{crit}})^{3/4}$},
where the initial critical energy density is given by~$\rho_{0}^{\textrm{crit}}\sim H_{\textrm{Inf}}^{2}M_{\textrm{Pl}}^{2}$.
Although~$\varphi$ starts with the value~$\varphi_{0}\sim H_{\textrm{Inf}}$,
it quickly reaches the Planck scale~$M_{\textrm{Pl}}$, therefore
we take the moderate value~$\varphi\sim\sqrt{H_{\textrm{Inf}}M_{\textrm{Pl}}}$
at early times so that~\begin{equation}
|\dot{\theta}|\sim10^{-8}\,\frac{(\rho_{0}^{\textrm{crit}})^{3/4}}{\varphi^{2}}\left(\frac{a}{a_{0}}\right)^{\!-3}<10^{-8}\sqrt{H_{\textrm{Inf}}M_{\textrm{Pl}}}.\end{equation}
 Even if we now choose the low value~$V_{0}\sim H_{\textrm{Inf}}^{3}M_{\textrm{Pl}}$
in the quintessence potential~$V(\varphi)=V_{0}e^{-\xi_{1}\varphi/M_{\textrm{Pl}}}$,
we obtain\begin{equation}
\frac{\dot{\theta}^{2}}{V''(\varphi)}<10^{-16}\frac{M_{\textrm{Pl}}^{2}}{\xi_{1}^{2}H_{\textrm{Inf}}^{2}}\sim10^{-4}\ll1,\end{equation}
 where we used~$H_{\textrm{Inf}}\sim10^{-7}\, M_{\textrm{Pl}}$ and~$\xi_{1}\sim10$.
At very late times in the cosmological evolution with a low Hubble
scale~$H\ll H_{\textrm{Inf}}$ we find~$V''(\varphi)=\mathcal{O}(H^{2})$
\cite{Steinhardt:1999nw},~$\varphi\gtrsim M_{\textrm{Pl}}$ and~$(a/a_{0})^{-3}=(H/H_{\textrm{Inf}})^{3/2}$
for a long period of radiation dominance. Thus, the quintessence field
is also stable in this epoch:\begin{equation}
\frac{\dot{\theta}^{2}}{V''(\varphi)}\sim10^{-16}\frac{H}{M_{\textrm{Pl}}}\ll1.\end{equation}
 We conclude that the quintessence scalar field in our scenario does
not seem to suffer from instabilities, at least within the context
of this simplified analysis. In principle, we could perform a similar
analysis also for the mediating field~$\chi$. However, since it
decays into neutrinos very quickly, a possible instability of~$\chi$
should not harm our scenario. Additionally, Q-Balls built from this
field could also decay into leptons \cite{Cohen:1986ct,Multamaki:1999an,Clark:2005blub}.
For a deeper discussion of Q-Balls see e.g.\  Ref.~\cite{Enqvist:2003gh}
and the references therein. 

In the last part of this section we want to motivate the statement
that particle processes only enter the equations of motion of the
fields in form of the decay term~$\Gamma\dot{\chi}$ in Eq.~(\ref{eomchi})
and make it plausible that this might even hold within a full quantum-mechanical
treatment of the quintessence field. However, one has to admit, that
such a treatment of the quintessence field would go far beyond the
scope of this paper and more subtleties would arise than the ones
mentioned here. 

\begin{figure}
\begin{center}\includegraphics[%
  clip,
  height=3.5cm,
  keepaspectratio]{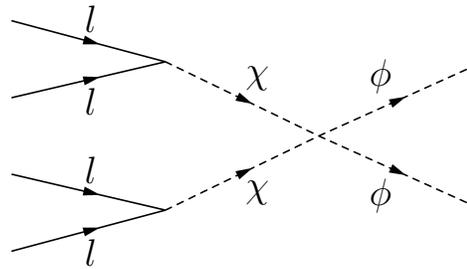}\vspace{-0.5cm}\end{center}

\caption{\label{wash-out}\textit{Example for an additional process between
$\phi$ and the leptonic sector.}}
\end{figure}

Still, we notice that one could also allow additional particle processes
like the one in Fig.~\ref{wash-out}, which might destabilize the
$\phi$ condensate.%
 However, even if this was done it is unlikely that they would have
a large impact on the scenario due to the following arguments. We
first notice that at the beginning of the scenario the masses of the
scalar particles (proportional to the second derivative of the corresponding
potential) are of the order of $H_{\textrm{Inf}}$ or smaller. By
dimensional analysis the reaction rates $\Gamma_{i}$ for the mentioned
processes should therefore be of the same order of magnitude at this
time (assuming coupling constants to be of order one). However, before
the corresponding time elapses the mass of $\chi$ has increased by
several orders of magnitude. Since these masses appear in the denominator
of the reaction rates the ratio $\Gamma_{i}/H$ will now always be
smaller than one and the reaction should never have enough time to
take place as also resonances do not seem to occur. 

We also note that additional decay channels of $\chi$ would not harm
our scenario as long as the lepton asymmetry is transferred to SM
particles before the electroweak phase transition. A decay channel
which could potentially be dangerous is a decay into hypothetical
quanta of the quintessence field, if they represent physical degrees
of freedom, via the couplings of the form $\lambda_{1,2}\phi^{2}\chi^{2}$.
However, a simple estimate for the decay rate $\Gamma_{2\chi\rightarrow2\phi}\approx\lambda_{1,2}^{2}|\chi|^{2}/(8\pi m_{\chi})$
\cite{Dolgov:1991fr} shows that it is both suppressed by the increase
of $m_{\chi}$ and the decrease of $|\chi|$ as compared to the rate
$\Gamma_{\chi\rightarrow2\nu}=g^{2}m_{\chi}/(8\pi)$ for the decay
into neutrinos.

\section{\label{sec:Conclusions}Summary and Conclusions}

In this work we have proposed a \BmL{} conserving baryogenesis scenario,
in which a \BmL{} asymmetry is hidden in the DE sector. This is achieved
by the introduction of two complex scalar fields carrying lepton number,
where one of them is responsible for DE in a quintessence-like manner
while the other one mediates between this field and the SM particles
(including right-handed neutrinos). An initial phase difference between
the VEVs of these two fields results in compensating but non-zero
lepton asymmetries for them. While the quintessence field has a tracking
behavior, the mediating field acquires a large mass and eventually
decays, hereby transferring its asymmetry to the fermionic sector.
Finally sphalerons partially transform this lepton asymmetry to a
baryon asymmetry. We quantitatively determined the resulting baryon
asymmetry numerically and also gave an analytic formula based on a
WKB approximation. Both results are consistent and show that the observed
value can easily be achieved. Furthermore, we have analyzed potential
sources of instability like Q-Ball formation and did not find any
hints for an unstable behavior. 

Of course, a deeper understanding of several topics is desirable.
A thorough knowledge about the VEVs of scalar fields after inflation,
especially in the case of complex potentials, could help to restrict
the parameter space of our model immensely. Also, as stated before,
a full quantum-mechanical treatment of the quintessence field is wish-full.
This could enable us to further restrict the couplings in our scenario,
while it might also make way for new baryogenesis models in which
the quintessence field is directly coupled to SM particles. 

The possible \BmL{} quantum numbers of the vacuum could also have
interesting consequences for additional couplings between DE and SM
fields or dark matter. Moreover quantum numbers of the quintessence
field could include hints at its origin. 

In conclusion, we presented a specific model that leads to the observed
baryon asymmetry of the universe with a strict conservation of~\BmL{}
and the storage of a corresponding asymmetry in the DE sector. 

\begin{acknowledgments}
We would like to thank M. Lindner as well as A. Anisimov, E. Babichev
and A. Vikman for useful comments and discussions. This work was supported
by the {}``Sonderforschungsbereich 375 f\"{u}r Astroteilchenphysik
der Deutschen Forschungsgemeinschaft''. F.~B. wishes to thank the
{}``Freistaat Bayern'' for a doctorate grant. MTE
wishes to thank J. Hamann for usefull comments and the {}``Graduiertenkolleg
1054'' of the {}``Deutsche Forschungsgemeinschaft'' for financial
support. 

\bibliographystyle{apsrev}
\bibliography{references-BAQuint}

\end{acknowledgments}

\end{document}